\documentclass[preprint,12pt]{article}

\pdfoutput=1

\usepackage{graphicx}
\usepackage{longtable}  
\usepackage{lscape}     
\usepackage{amssymb,amsfonts,amsmath}
\usepackage{color}
\usepackage{algorithm}
\usepackage{algorithmic}

\def\bm#1{\mbox{\boldmath$#1$}}  
\newcommand{\flx}[1]{f_{\bm{\Lambda}_{#1}}\left(\bm{x}\right)}
\newcommand{\flxx}[1]{f_{\bm{\Lambda}{#1}}\left(\bm{x}\right)}
\newcommand{\rholx}[1]{\rho_{\bm{\Lambda}_{#1}}\left(\bm{x}\right)}
\newcommand{\rholxx}[1]{\rho_{\bm{\Lambda}{#1}}\left(\bm{x}\right)}
\newcommand{\bL}{\bm{\Lambda}}
\newcommand{\htti}{\hat{t}}
\algsetup{indent=1em}

\title{Multidimensional integration through Markovian sampling 
under steered function morphing: a physical guise from statistical mechanics}

\date{\vspace{-5ex}}

\begin{document}

\maketitle

\begin{center}
\author{Mirco Zerbetto}
\author{Diego Frezzato*}
\\*[1cm]
Dipartimento di Scienze Chimiche, Universit\`a degli Studi di Padova, 
via Marzolo 1, I-35131, Padova, Italy
\\*[1cm]
* diego.frezzato@unipd.it
\end{center}

%

\begin{abstract}
We present a computational strategy for the evaluation of multidimensional integrals on
hyper-rectangles based on Markovian stochastic exploration of the integration domain 
while the integrand is being morphed by starting from an initial appropriate profile. 
Thanks to an abstract reformulation of Jarzynski's equality applied in stochastic thermodynamics to evaluate the free-energy profiles along 
selected reaction coordinates via non-equilibrium transformations, it is 
possible to cast the original integral into the exponential average of the distribution 
of the pseudo-work (that we may term "computational work") involved in doing the function 
morphing, which is straightforwardly solved.
Several tests illustrate the basic implementation of the idea, and show its performance in terms of computational time, accuracy 
and precision. The formulation for integrand functions with zeros and possible sign changes is also presented. 
It will be stressed that our usage of Jarzynski's equality shares similarities
with a practice already known in statistics as 
Annealed Importance Sampling (AIS), when applied to computation of the normalizing constants of
distributions. In a sense, here we dress the AIS with its "physical" counterpart borrowed 
from statistical mechanics. 
\end{abstract}
%
%

\
%
\section{Introduction and motivation}\label{sec1}
Multivariate numerical integration on hyper-rectangles of large dimension $N$ is often encountered in many areas
of engineering and of physical, natural and social sciences. It is known 
that the accuracy of polynomial approximation techniques (e.g. ''cubature rules'', generally speaking) exponentially
degrades on increasing the number of variables, so that beyond $N \sim 10$ these routes have to be abandoned. In these 
situations one is forced to resort, at last, to statistical integration techniques 
framed in the broad family of Monte Carlo (MC) routes.
From the 40's of past century, when statistical sampling was conceived at Los Alamos Laboratory by scientists like von Neumann, 
Ulam and Metropolis to face practical problems in particle physics \cite{metropolis49}, several improvements have been done to 
accelerate convergence, reduce uncertainty on the estimate, and limit the degradation of the efficiency at fixed 
computational cost as $N$ increases. In particular, we assisted at a flowering of variants of the basic 
''sample mean integration'' (e.g. Quasi-Monte Carlo \cite{schuerer03, cools07}), and of the classical Metropolis-Hastings 
''importance sampling'' MC \cite{metropolis53,Hastings70} (IS-MC in the following) which is widespread in 
many branches of molecular and condensed matter physics where configurational partition functions need to be
calculated. For example, functions of the Genz's testing-family \cite{genz84} can by routinely integrated with high accuracy 
and low computational cost up to hundreds of variables by means of optimized 
quasi-random sampling strategies \cite{cools07}. Despite of the large efforts that have been done to improve statistical integration, 
and of the constantly increasing computational power at disposal, there is still need for efficient algorithms which are robust 
regardless of the peculiar kind of integrand function.

In this work we present a tool to perform multivariate integration by exploiting a Markovian 
stochastic exploration of the integration domain while the integrand function is morphed in a controlled 
(''deterministic'') way. The strategy is rooted in an abstract interpretation of Jarzynski's equality (JE in
the following), which was derived about fifteen years ago in the context of the thermodynamics of systems
(mainly macromolecular) subjected to thermal fluctuations and driven out of equilibrium by an external
mean which has full control on a selected set of structural parameters \cite{JAR5018}. 
Since a detailed description of the JE on physical grounds would be not pertinent to the spirit of the present
communication, we address the interested reader to the excellent reviews of refs. 
\cite{JAR331, JAR329, POH10235} which comprise theory and experiments. We shall give here
below only the essential lines 
to appreciate the integration methodology that we are going to present. 

In the essence, if $A$ denotes the Helmholtz free-energy of a system at thermal equilibrium
and {\em constrained} in a certain state specified by a set of controllable parameters, say $\bL$, 
the free-energy change from state "1" to state "2" is $\Delta A_{1 \to 2} = \ln (Z_2 / Z_1)$ 
where $Z_1$ and $Z_2$ are the so-called canonical configurational partition functions \cite{huang}. Explicitly, these 
are the integrals $Z_{1(2)} = \int d \bm{x} \, e^{-V_{1(2)}(\bm{x})/k_BT}$ made over all {\em unconstrained} 
structural variables $\bm{x}$ of the system, which fluctuate due to the contact with the thermal 
bath. $V_{1(2)}(\bm{x})$ is the configuration-dependent energy of the system
(here for two states "1" and "2" corresponding to $\bL_1$ and
$\bL_2$, respectively), $T$ is the absolute temperature, and $k_B$ the Boltzmann constant. The JE allows
to evaluate $\Delta A_{1 \to 2}$ by avoiding the explicit calculation of the partition functions, which
becomes highly demanding (or even unfeasible) as the number of variables grows. Namely, 
the JE states that the free-energy difference can be cast into the exponential average of the amount of work 
$w_{1 \to 2}$ performed by an external mean to drive the system from the equilibrium state "1" to the 
state "2" along non-equilibrium transformations where the controlled state-parameters are changed 
according to a prescribed (but arbitrarily chosen) protocol $\bL(t)$ of finite duration. 
Explicitly, the JE is $\Delta A_{1 \to 2} = - k_B T \ln \langle e^{-w_{1 \to 2}/k_BT} \rangle$, where 
the average is taken over the ensemble of {\em stochastic} trajectories $\bm{x}(t)$, each generated while 
the state-parameters are deterministically changed. We stress that, in thermodynamics, the work is 
identified with the amount of energy exchanged between the external mean and the system through a detailed 
action on some degrees of freedom of the system (the set $\bL$ in this case); thus, $w_{1 \to 2}$ 
is obtained by accumulating the infinitesimal contributions 
$\delta t \times \left. \partial V_{\bL(t)}(\bm{x})/ \partial t 
\right|_{\bm{x} = \bm{x}(t)}$ to be evaluated (i.e., measured in experiments by knowing the
applied forces, or calculated if trajectories are simulated) at each instant along the specific
trajectory. The JE is valid under the mild
conditions of {\em i}) fluctuations of the uncontrolled variables is a 
Markovian (memory-less) process \cite{gardiner}, and {\em ii}) the 
trajectory would sample the underlying canonical distribution proportional to
$\exp\left\{-V_{\bL^\ast}(\bm{x}) /k_B T\right\}$ after the protocol was stopped at some state $\bL^\ast$. 
The JE has been extensively applied to construct entire 
free-energy profiles between states connected by real, simulated, or even artificial steered trasformations. 
Typical examples are real mechanical unfolding/refolding of biopolymers (e.g., RNA hairpins) by means of laser 
tweezers \cite{Liphardt02},
simulated detachment of chemicals from binding sites of proteins \cite{Nicolini13}, and virtual "alchemical" transformations of
molecular moieties to evaluate solvation free-energies in a given environment \cite{Cossins09}. 
The strength of the JE is that an accurate estimate of $\Delta A_{1 \to 2}$ (which means an accurate
estimate of the ratio $Z_2/Z_1$) could be achieved from a limited number of runs starting from initial 
configurations $\bm{x}(0)$ drawn from the pool belonging to the same equilibrium state, and employing always the same protocol. 
In the experimental context, the JE offers the remarkable link between a measurable quantity (the work) and the change 
of free-energy for a nanoscale system. On computational grounds, the effort required 
by the JE machinery to compute the ratio $Z_2 / Z_1$ from a set of simulated transfomations is usually much lower than that required 
to compute the single partition functions by direct integration, and the accuracy of the 
outcome remains acceptable even when standard routes, like the Metropolis IS-MC, fail.
\cite{notex1}

In the present work we "borrow" the JE and adopt it out-of-context to efficiently solve 
multidimensional integrals, just exploiting (by analogy with physical transformations) the possibility to 
substitute the explicit integration with the evaluation of what we can term the "computational work" in doing 
the "externally controlled" build-up of the integrand function (that we shall term "morphing" 
in the following) starting from an easily integrable known profile.
By viewing the canonical configurational partition functions of initial and final states (physical context) as nothing but multidimensional 
integrals of positive-valued functions (computational context), and the steered non-equilibrium transformations 
on few system parameters while all the remaining, uncontrolled, degrees of freedom continue to fluctuate (physical context) as the analogous 
of the morphing of the integrand {\em while} the integration domain is stochastically explored by Markovian moves compatible
with the actual function landscape itself (computational context), then an adaptation/exploitation of the JE is straight devised 
as a tool to perform multidimensional integration.
In a na\"ive picture of adaptive nature of Markov dynamics over an evolving landscape, sampling the integration 
domain {\em while} the features of the integrand function ''grow'' following an arbitrarily prescribed protocol 
is more efficient than exploring directly the given integrand landscape. 

To the best of our knowledge, all applications of the JE are strictly pertinent
to the original physical contexts (mainly chemical, namely the estimation of mean-field-potentials along specific coordinates of 
complex molecular or supra-molecular systems). With the present contribution we intend to enucleate from the JE,
taking out the physical traits, its essential computational feature of ''machinery to perform multidimensional integration'' in efficient way.
Although apparently trivial once formulated, we believe that our abstract rephrasing may disclose the real powerful
of the JE and bring it to a broader audience than the community of researchers active in 
physical-chemical areas. Here we stress the crucial point that in statistics, namely in practices to sample 
from general distributions, the strategy named Annealed Importance Sampling (AIS) due to Neal \cite{Neal01} 
(see also refs. \cite{Creal12} and \cite{Tokdar09} for recent reviews)
leads to an analogous of the JE. Being targeted to the sampling from a "complicate" (eg., multimodal) distribution, 
the AIS consists in morphing an initial easy-to-sample distribution up to the target one by filling 
the gap with a freely chosen number of bridging distributions. Surprisingly, the AIS seems to be almost unknown in the
physical-chemical community, apart of few exceptions \cite{Lyman07}.
By following Neal's work, one can find a direct matching
with Jarzynski's relation when the developing of the intermediate distributions is translated in terms of transformation 
protocol; on the other way around, the JE framework gives to the AIS a "physical guise" where the concept of "work" is the key-feature. 
Once such a connection is made, our feeling (and auspice) is that the huge amount of expertises achieved in latter decade
in the physical-chemical area can be trasferred to the development of pure numerical methods of stochastic integration. 
Suggestions will be given in
the course of our exposition, namely {\em i)} the possibility to generate paths ${\bf x}(t)$ via stochastic differential equations
(Langevin-like equations) in place of Monte Carlo chains, and hence to be inspired by the
huge amount of studies of molecular Brownian-like dynamics in condensed 
fluid phases; {\em ii)} devise optimal morphing protocols to improve accuracy and precision
of the integral estimate, in analogy with what is done in molecular practices
(both simulations and experiments)
to reduce the energy dissipation during a steered transformation; {\em iii)} take benefit 
from the good practices, developed for the free-energy-difference evaluations, to control/estimate the errors on the outcome; 
{\em iv)} be aware of the vast physico-chemical 
literature of the last decade, where smart improvements of the basic implementation of the JE  
are proposed (the interested reader can find a valuable overview in ref. \cite{Ytreberg06}, and specific 
indications will be given in the section "Outlines and perspectives").

In developing our exposition we shall present a simple mean to perform the function morphing, namely the rising of
the whole integrand function from an initial flat profile \cite{notex2}.
In addition, with a simple trick we shall leave the context of positive-valued
functions to account for general profiles of the integrands.
Besides of drawing the main methodological lines, we also developed the JEMDI (Jarzynski Equality 
MultiDimensional Integration) library, an optimized and easy to use C++ 
algorithm implementing our approach to stochastic integration, which is freely distributed as open-source 
software for tests and further developments \cite{JEMDI}. 

The article continues as follows. First we provide a general outline about 
how to frame Jarzynski's equality in the abstract context of integration. Then we opt for the morphing
protocol from a flat profile, which is the most safer and case-independent choice if the details of the
integrand are unknown. In section "Computational issues" we briefly present the JEMDI software which
currently implements such a choice; details of the numerical solver are provided 
in the Supplementary Material. Then we test the algorithm on model cases, and give the proof of its outperforming efficiency 
when compared to the standard IS-MC route (i.e., the direct evaluation of the integral without morphing). 
This is not surprising for experts in JE (or AIS), since non-equilibrium routes are known to achieve a 
likely result in a reasonable computational time while such a standard counterpart completely fails. 
Furthermore we provide an estimate of the uncertainty and a
criterion to judge the reliability of the outcome. In our tests, exploration of the integration domain will be made mostly by
means of IS-MC moves, but we give also an example where Langevin dynamics are employed. The final section is 
devoted to remarks and perspectives.

\section{Stochastic integration under controlled morphing}\label{sec2}
In this section we shall present the integration strategy rooted in the abstract formulation/usage of Jarzynski's equality. 
We start by considering the physically framed case of positive-valued integrand functions. 
 Then we extend the method to general integrands that may have zeros and/or sign changes within the integration domain.
Throughout the text we make free use of physical terminology by leaving to the reader the effort to keep any 
step at the most abstract level.

\subsection{Abstract formulation of Jarzynski's equality}\label{sec2sub1}
Let us first consider the case of a real and positive-valued function $\flx{}$, with 
$\bL$ a set of parameters defining its profile and $\bm{x} \in I$ the argument as a $N$-dimensional array of real 
variables. The function must be bounded and continuous in the interior of the integration domain 
$I$ with the further requirement that at boundaries the primitive function has a finite limit if $\flx{}$
diverges (see discussion in section \ref{sec4}). Our purpose is to set up an efficient route to determine the integral
\begin{equation}\label{eq_general_integral}
E\left(\bL\right)=\int_{I}d\bm{x}\flx{}
\end{equation}
If the integral is known for a certain set of parameters, say $\bL_0$, one can write
\begin{equation}\label{eq_morphed_integral}
E\left(\bL\right) = E\left(\bL_0\right)\Phi\left(\bL,\bL_0\right)
\end{equation}
where $\Phi\left(\bL,\bL_0\right)$ is the ''morphing factor'', related to the change of $\flx{0}$ into $\flx{}$, 
to be determined.

Let us imagine to randomly pick a point $\bm{x} \in I$ from the distribution $\rholx{0} \propto \flx{0}$, and drive 
the morphing in a deterministic way according to an arbitrarily chosen protocol $\bL(\htti)$  where $\htti$
is treated as ''time'' variable varying from zero to one (as a matter of fact, it is nothing but a dimensionless
progression variable); "during" such a transformation, be $\bm{x}$ free to explore the integration 
domain by following a general type of stochastic Markov ''dynamics'' \cite{gardiner},
$\bm{x}(\htti)\rightarrow\bm{x}(\htti+\delta\htti)$, where $\delta\htti$ is the propagation step. By producing 
a large number of these transformations each conducted by employing the same protocol $\bL(\htti)$, starting from 
different initial states all sampled from the distribution $\rholx{0}$, a non-equilibrium
distribution $\rholxx{(\htti)}$ will develop. The requirement of Markov dynamics is important 
to assure that if the morphing was stopped at a certain $\htti^\star$ and the dynamics continued, the distribution would 
relax to the underlying target distribution proportional to $\flxx{(\htti^\star)}$, that is 
$\lim\limits_{\htti\to\infty} \rholxx{(\htti)} \propto\flxx{(\htti^\star)}$. 
Consider now the following path-integral along the $i$-th stochastic trajectory $\bm{x}(\htti)_{tr,i}$,
\begin{equation}\label{eq_work}
w_i=\int_0^1 d\htti \, \dot{\bL}(\htti)\cdot
\left.\frac{\partial u\left(\bm{x} (\htti)_{tr,i},\bL\right)}{\partial \bL}\right|_{\bL=\bL(\htti)}
\end{equation}
where we have introduced
\begin{equation}\label{eq_pseudo-potential}
u(\bm{x},\bL)  = -\ln\left(\flx{}\right)
\end{equation}
We can state that the morphing factor in Eq. \eqref{eq_morphed_integral} can be expressed by the following 
limit taken over an infinitely large number $N_{tr}$ of trajectories:
\begin{equation}\label{eq_morphing_factor}
\Phi\left(\bL,\bL_0\right)=\lim\limits_{N_{tr}\to\infty}\frac{1}{N_{tr}}\sum_{i=1}^{N_{tr}}e^{-w_i}
\end{equation}
The proof simply rests on the direct recognition that Eq. \eqref{eq_morphed_integral} corresponds to the JE 
if the following identifications are done: \emph{i}) $E\left(\bL_0\right)$ and $E\left(\bL\right)$ are homologous 
to the configurational partition functions evaluated over the canonical distributions 
respectively originated by the pseudo-potentials $u(\bm{x},\bL_0)=-\ln\left(\flx{0}\right)$ and 
$u(\bm{x},\bL)=-\ln\left(\flx{}\right)$; \emph{ii}) the morphing of the function corresponds to a steered 
transformation following the deterministic protocol $\bL (\htti)$; \emph{iii}) the Markovian exploration 
of the domain is equivalent to the Markovian dynamics over the uncontrolled degrees of freedom in the physical 
context of non-equilibrium transformations; \emph{iv}) the path-integral in Eq. \eqref{eq_work} is interpreted 
as the ''work'' to morph the function while the integration domain is stochastically sampled; 
\emph{v}) Eq. \eqref{eq_morphing_factor} is the analogous of the work-exponential-average in the proper JE, 
being $-\ln\left(\Phi\left(\bL,\bL_0\right)\right)$ equivalent to the free-energy difference between 
the morphed and the initial states. 
Such a one-to-one mapping of our computational problem into the physical context allows us to take
{\em directly} Eq. \eqref{eq_morphed_integral} with Eqs. \eqref{eq_work}-\eqref{eq_morphing_factor} as an
{\em exact} result right on the basis of the sound validity (as theorem) of the JE itself.
The interested reader may find a transparent derivation of the JE, for example, in Section II 
of ref. \cite{VAI054107}. Looking at such a derivation, the tight connection with Neal's AIS method will
appear (see sec. 2 of ref. \cite{Neal01}); namely, by taking a continuum of bridging distributions each
labeled by $\bL (\hat{t})$, eq. 5 of ref. \cite{Neal01} gives exactly the factors $\exp (-w_i)$, and 
eq. 2 yields (for the specific application) the JE estimator on a finite number 
of transformations.  

The above formulation is rather general. Notice that, unlike the original physical context where the energetics 
of the system and the dynamical responses are often fixed by the nature of the sample, here there 
is plenty of room to choose the reference state $\bL_0$ for which $E(\bL_0)$ is known, to optimize the protocol 
$\bL(\htti)$ (meaning that also single parameters $\lambda_1(\htti)$, $\lambda_2(\htti)$, $\dots$ could be varied 
independently and in different ways),
and to choose/optimize the kind of Markov exploration of the integration domain. The target is to achieve 
numerical convergence on $\Phi(\bL,\bL_0)$ with the lowest number of trajectories $N_{tr}$ of shortest
length (i.e., number of elemental propagation steps, see below). Insights will be given in what follows.

\subsection{Stochastic exploration of the integration domain}\label{sec2sub2}
The choice of the evolution law $\bm{x}(\htti)\rightarrow\bm{x}(\htti
+\delta\htti)$ is guided, in the computational context,
only by the need of generating a Markov chain which ensures that if the morphing is stopped then the 
dynamics would settle over the stationary ''equilibrium'' distribution as discussed above. {\em Any} propagator able to create such
a Markov chain can be employed in the algorithm. Regardless of the specific kind of evolution law, reflecting 
conditions have to be applied at the boundaries of $I$.

A straightforward method to create a Markov chain is to explore the integration domain by means of IS-MC moves 
of maximum length $\delta_{{\rm max},i}$ in each dimension $i$ \cite{metropolis53, ALLENTILDESLEY}. 
In practice, the trajectory is broken into $N_{\rm steps}$ steps of equal "duration" $\delta \hat{t} = 1/{N_{\rm steps}}$.
A generic step $1 \leq s \leq N_{\rm steps}$ begins at $\hat{t}_{s-1} = (s-1)\delta \hat{t}$ and ends 
at $\hat{t}_s = s\delta \hat{t}$, and
consists of a first part where the function morphing proceedes for the duration $\delta \hat{t}$ 
at the location $\bm{x}(\hat{t}_{s-1})$ frozen, and a second part where a move is meant to happen 
instantaneously up to $\bm{x}(\hat{t}_s)$. In schematic form we have 
$\bm{x}(\hat{t}_{s-1}) \stackrel{\rm morph.}{\rightarrow} \bm{x}(\hat{t}_{s-1}) 
\stackrel{\rm move}{\rightarrow} \bm{x}(\hat{t}_s)$.
An unbiased move is first attemped with the sole limitation that
$|x_i(\hat{t}_s)-x_i(\hat{t}_{s-1})| \leq \delta_{{\rm max},i}$. The move is readily
accepted if it is downhill, i.e., if $u(\bm{x}(\hat{t}_s),\bL(\htti_{s})) \leq u(\bm{x}(\hat{t}_{s-1}),\bL(\htti_{s}))$. 
On the contrary, a random number $\alpha$ is uniformly generated between 0 and 1 and the move is accepted if 
$\exp\{u(\bm{x}(\hat{t}_s),\bL(\htti_{s})) -u(\bm{x}(\hat{t}_{s-1}),\bL(\htti_{s}))\} < \alpha$, rejected otherwise. 
With such a classical scheme due to Metropolis {\em et al.} \cite{metropolis53}, the requirement of relaxation to the underlying
canonical distribution after stopping the morphing is automatically fulfilled by construction.
Work is performed only in the morphing part of a step, thus
$\delta w(s) = \delta\htti \, \dot{\bL}(\htti_{s-1})\cdot
\left. \partial u\left(\bm{x},\bL\right)/\partial \bL \right|_{\bm{x}=\bm{x}(\htti_{s-1}),\bL=\bL(\htti_{s-1})}$. The
global amount per trajectory is then obtained by summing over the steps, which corresponds to take the discretized
formulation of the integral in Eq. \eqref{eq_work} along the path. 
The IS-MC scheme is a fast and simple procedure that enables long-range rapid exploration and only requires the
evaluation of the integrand function at each step.

Alternatively, evolution laws based on stochastic 
differential equations can be adopted, which require to supply also first-order derivatives 
(which must be bounded and continuous in $I$) of the integrand function. The simplest and physically-framed propagation 
scheme of such a kind makes use of a Langevin-like equation corresponding to a Brownian-like exploration of the 
integration domain \cite{gardiner}, i.e.
\begin{eqnarray}\label{eq_langevin}
\nonumber
\bm{x}(\htti+\delta\htti) &=&\bm{x}(\htti) + \delta\htti
\Bigg[\left(
\frac{\partial}{\partial\bm{x}}
\cdot \bm{D}(\bm{x},\bL)\right)^{T}+\\
&&
-\bm{D}(\bm{x},\bL)
\frac{\partial u(\bm{x},\bL)}{\partial \bm{x}}
+ \, \sqrt{2} \, \bm{D}^{1/2}(\bm{x},\bL) \, \bm{\eta}(\htti)
\Bigg]_ {\left \lvert
\begin{array}{l}
\scriptsize{\bm{x}=\bm{x}(\htti)} \cr
\scriptsize{\bL=\bL(\htti)}
\end{array}
\right.}
\end{eqnarray}
where $\bm{\eta}(\htti)$ is the vector whose
$N$ entries are independent sources of white noise: $\langle\bm{\eta}(\htti)\rangle=\bm{0}$ and 
$\langle\bm{\eta}(\htti)\otimes\bm{\eta}(\htti')\rangle=\delta(\htti-\htti')\bm{1}$, with $\bm{1}$ the
$N\times N$ identity matrix, $\delta(\cdot)$ the Dirac's delta-function, and  $\langle \dots \rangle$ are
ensemble averages over the distribution of noise magnitudes. In practice, at the actual
$\htti$, for each variable $x_j$ one generates $\eta_j(\htti)=s(0,1)/\sqrt{\delta\htti}$ where $s(0,1)$
is a value randomly sampled from a distribution (usually Gaussian, but not necessarily so, see remarks 
in the Supplementary Material) with zero mean and unit variance. Finally, in Eq. \eqref{eq_langevin}
$\bm{D}(\bm{x},\bL)$ is a "diffusion matrix" which can be freely designed to be both point-dependent and 
deterministically modulated along the morphing, under condition that it must be real-valued, symmetric and 
positive-definite. This freedom can be exploited, in principle, to optimize in subtle way the stochastic 
exploration of the integration domain. 
As for the IS-MC evolution, a single step of duration $\delta \hat{t}$ is meant to be constituted by a 
morphing part followed by a Langevin propagation. The infinitesimal amount of work is evaluated exactly as for
the IS-MC case, and the work per trajetory follows by summing all contributions.

\subsection{Choice of the reference state}\label{sec2sub3}
The choice of the reference state $\flx{0}$ is crucial to improve the efficiency of integration. 
A good balance should be found between closeness of $\flx{0}$ to $\flx{}$ 
(intuitively this would reduce the amount of ''dissipation'' in analogy with the physical steered transformations),
simplicity of its integration (possibly analytical) to get $E(\bL_0)$, and capability to sample  
initial configurations quickly and without artefacts from the distribution $\propto \flx{0}$
(such a difficulty is lowered if a factorization is recognized in the integrand function). 
Clearly, only a well educated guess may lead to identify a suitable integrable function $\flx{0}$ of which 
$\flx{}$ is intended to be a perturbed form. On the contrary, a dangerous bias might be introduced. 

In the absence of some {\em a priori} knowledge, the simplest and safer choice is to start from a flat profile of
the associated potential, that is $u(\bm{x},\bL_0)=c$ for all $\bm{x}$, so that $E(\bL_0)=V_Ie^{-c}$ being $V_I=\int_I d\bm{x}$ the 
volume of the integration domain. In this case the initial sampling reduces to an unbiased random drawing of 
independent points in the $N$-dimensional space. 
A proper value of $c$ can be set as $c=\ln\left(V_I/E_{trial}\right)$ where $E_{trial}$ is a guess (even very rough) of 
the integral; with such a choice the morphing factor is expected to fall close to one.

\subsection{Best estimate of the integral and related uncertainty}\label{sec2sub4}
Assessment of the outcome reliability on statistical grounds takes benefit of the sound experience gained in the
context of the JE applied to free-energy-difference calculations in physical (mainly molecular) systems.
Having generated $N_{tr}$ trajectories, the best estimate of the integral is
\begin{eqnarray}\label{eq_estJARZ}
E_{N_{tr}}(\bL)=E(\bL_0) \, \Phi_{N_{tr}} \;\;\; , \;\;\;
\Phi_{N_{tr}} =\frac{1}{N_{tr}} \sum_{i=1}^{N_{tr}} e^{-w_i}
\end{eqnarray}
For $N_{tr} \to \infty$ such a relation is {\em exact}, while for any finite number of transformations
the estimate bears an error, $\delta(N_{tr})=E_{N_{tr}}(\bL)-E^\ast(\bL)$ being
$E^\ast(\bL)$ the true value, which is due to the limited sampling of the low-work "wing" of the 
distribution of work values, $p(w)$. By looking at Eq. \eqref{eq_estJARZ}, 
important trajectories which largely contribute are those with lower values of $w_i$; 
on the other hand, these trajectories are rarely encountered (because of the low values $p(w_i)$ in the distribution 
wing), hence their frequency of appearance, in a statistical sample of finite size, may largely deviate 
from the actual probability. In particular, the distribution of $\delta(N_{tr})$ (by supposing to
repeat infinite times the calculation with a set of $N_{tr}$ trajectories) displays an average shift
$\overline{\delta E}$ , that is a $N_{tr}$-dependent systematic error, plus a broadening 
which arises from an intricate interplay between the (unknown) features of the work-distribution-function 
and the finiteness of the ensemble of work values at disposal. 

An indication about the systematic error can be inferred from the Zuckerman - Woolf theory developed 
for free-energy-difference calculations \cite{ZUC180602}. It was already known that, 
{\em on average}, one gets an overestimation (see eq. 56 of ref. \cite{JAR5018}), but the 
authors were able to derive a "universal
relation" (valid under mild conditions) which links the systematic error to the variance 
of the outcomes. Turning to our context, if we propagate back such an uncertainty to the integral (we recall
that the morphing factor is equivalent, in the essence, to the exponential of minus 
a free-energy-difference), the result for $N_{tr}$ sufficiently high is that the integral
is {\em on average} underestimated by  
$\overline{\delta E} \simeq -\sigma_E^2 /2E$ where $\sigma_E$ is the standard deviation 
of the distribution of outocomes $E_{N_{tr}}(\bL)$ (compare with eq. 17 of ref. \cite{ZUC180602}). 
This relation tells us that if $\sigma_E / E \ll 1$, then 
$\overline{\delta E} \ll \sigma_E$ (accurate outcome) and $\pm \sigma_E$ (precision
estimator) can be taken as a likely estimate of the interval of confidence. In the following we shall
indicate with $\delta E_{\rm stat} \equiv \sigma_E$ such an uncertainty.

At first instance one may evaluate $\sigma_E$ from the raw outcomes as
\begin{equation*}
\sigma_E \simeq E(\bL_0)  \left[ N_{tr}^{-1} (N_{tr}-1)^{-1} \sum_{i=1}^{N_{tr}} 
\left( e^{-w_i} - \Phi_{N_{tr}} \right)^2 \right]^{1/2}
\end{equation*} 
where $\Phi_{N_{tr}}$ is seen as average over $N_{tr}$ entries $\exp(-w_i)$. The
estimate can be eventually improved, for example, by means of resampling procedures from the dataset 
at disposal, such as the "bootstrap" route.
Here we pursue a different choice 
borrowed from the practice of "blocks averages" in free-energy-difference calculations. It consists in 
randomly splitting the whole dataset with  $N_{tr}$ entries into $M$ groups (in our tests, 
these groups are formed by $N_{tr,b} = N_{tr}/M$ consecutive trajectories as they are generated), 
and then taking
\begin{eqnarray}\label{eq_sigmaE}
\sigma_E \simeq E(\bL_0) 
\left[
M^{-1}(M-1)^{-1} \sum_{k=1}^{M} \left( 
\Phi_{N_{tr,b}, {\rm block} \, k} - \Phi_{N_{tr}}
\right)^2 
\right]^{1/2}
\end{eqnarray}
where the morphing factors $\Phi_{N_{tr,b}, {\rm block} \, k}$ are computed with the $N_{tr,b}$ 
trajectories of each $k$-th block. Eq. \eqref{eq_sigmaE}
is the estimate of the standard deviation of the mean, evaluated from the spread of $M$ partial outcomes; the formula 
is meant to be improved via t-Student correction for low values of $M$. Clearly, the result 
depends on the choice of $M$, although such a dependence is weak when $M$ is of the order of few tens.
The idea is to choose $M$ yielding blocks which are supposed to be large enough to provide 
"sensible" estimates of the integral \cite{notex3}.

We stress that the evaluation of $\sigma_E$ from the data at disposal can be 
highly inaccurate due to the poor sampling of $p(w)$, so that the estimated ratio 
$\sigma_E / E$ could result (apparently) small even if the systematic error is relevant. 
Therefore one needs independent criteria to establish 
if $p(w)$ is well sampled to allow one to take $\sigma_E / E \ll 1$ as reliable indicator of
accurate integration. 
A criterion has been provided by Jarzynski (see note 23 of ref. \cite{JAR5018}), 
Hummer (section IV of ref. \cite{Hummer01}), and reaffirmed by others \cite{Oberhofer05, Ytreberg06}: 
good sampling of the low-work wing of $p(w)$  is likely attained if $\sigma_w \leq 1$, where $\sigma_w$  
is the standard deviation of the work values (still to be estimated, unfortunately, from the finite
set of data at disposal). 
If such a condition is not fulfilled, one may slower the transformation protocol and/or increase 
$N_{tr}$. Here we pursue a different empiric criterion which is applicable when the IS-MC route is 
employed to make the Markov exploration of the integration domain (in section \ref{sec4} we shall
validate its effectiveness). Namely we evaluate the average percentage of accepted moves 
over the whole ensemble of paths and over the whole morphing schedule, $\overline{\%_{acc}}$.
Then by borrowing the recommendation for standard Metropolis MC practices \cite{ALLENTILDESLEY}, we simply 
check if $\overline{\%_{acc}}$ is around 50\%. This follows by the guess that an efficient sampling of the
integration domain (although on average) is associated to a slow-enough transformation (a "quasi-static" one
in the thermodynamics language), hence to low "dissipation" (energy dissipation in the thermodynamics 
acceptation, see remarks in section \ref{sec5}), and hence to a precise/accurate outcome.

\subsection{Basic implementation: homogeneous morphing at constant rate}\label{sec2sub5}
Here we shall tailor the above general scheme to the case of homogeneous morphing controlled by a unique 
parameter $\lambda(\htti)$ varying from $\lambda(0)=0$ to $\lambda(1)=1$. Moreover, the basic constant-rate 
''linear schedule'' is employed: $\lambda(\htti)=\htti$. Being interested in evaluating the integral
$\int_I d\bm{x}f(\bm{x})$, we set $f_0(\bm{x})=e^{-u_0(\bm{x})}$, $f_1(\bm{x}) =e^{-u_1(\bm{x})} \equiv f(\bm{x})$, and
$f_{\htti}(\bm{x})=e^{-[\htti u_1(\bm{x}) + (1-\htti)u_0(\bm{x})]}$ for the intermediate states. The path-dependent 
''morphing work'', Eq. \eqref{eq_work}, reads
\begin{equation}\label{eq_work_simple-protocol}
w_i=\int_0^1 d\htti \left[u_1\left(\bm{x}(\htti)_{tr,i}\right)-u_0\left(\bm{x}(\htti)_{tr,i}\right)\right]
\end{equation}
The special case of morphing from an initial flat profile is immediately obtained with
$u_0(\bm{x})=c$ in Eq. \eqref{eq_work_simple-protocol}.

\subsection{Extension to functions with zeros and sign changes}\label{sec2sub6}
The strategy sketched above directly follows from the abstract interpretation of the JE; strictly speaking, 
integrands must be positive-valued in order to be assimilated to Boltzmann factors. This severe ''physical'' 
limitation can be overtaken by means of a simple trick, namely splitting the general function $f(\bm{x})$ 
(which may have zeros and possibly change sign within the integration domain) as
\begin{equation}\label{eq_function_splitting}
f\left(\bm{x}\right)=f_{+}(\bm{x}) - f_{-}(\bm{x})
\end{equation}
where the components $f_{+}(\bm{x})$ and $f_{-}(\bm{x})$ are freely chosen positive-valued functions. 
The simplest choice is to apply
\begin{eqnarray}\label{eq_a_and_b}
f_{\pm}(\bm{x}) = \left[ \Phi(f\left(\bm{x}\right)) \pm f \left(\bm{x}\right)\right]/2
\end{eqnarray}
where $\Phi(\cdot)$ is any function which satisfies $\Phi(f)>|f|$. This ensures that,
by construction, both components are positive-valued and the integration strategy above presented can be applied separately 
to them. The global uncertainty on the net integral is then evaluated by summing in quadrature the two uncertainties.

In the present implementation we adopt the following form
\begin{equation}\label{eq_function_Phi}
\Phi(f) = K \sqrt{f^2 + \epsilon^2}
\end{equation}
with $K \geq 1$ and $\epsilon >0$ a small number suitably chosen (see below). One easily verifies that 
$f_{+}(\bm{x}) \simeq (K+1) f(\bm{x})/2$ and $f_{-}(\bm{x}) \simeq (K-1)f(\bm{x})/2 $ if
$f(\bm{x})>> \epsilon$, while $f_{+}(\bm{x}) \simeq (K-1) |f(\bm{x})| /2$ and 
$f_{-}(\bm{x}) \simeq (K+1) |f(\bm{x})| /2$ if
$f(\bm{x})<<-\epsilon$. Thus, for $\epsilon$ sufficiently small, where $|f| \gg \epsilon$ the two components in 
Eq. \eqref{eq_function_splitting} well approximate the original function (in absolute value) multiplied by positive factors 
$(K-1)/2$ or $(K+1)/2$. In the opposite limit $|f| \ll \epsilon$ one has that 
$f_{+}(\bm{x})\simeq f_{-}(\bm{x})\simeq K \epsilon /2$. 

A special situation is for $K = 1$. In this case, for
very small $\epsilon$ the component $f_{+}(\bm{x})$ selects the positive part of $f(\bm{x})$ while 
$f_{-}(\bm{x})$ selects the negative part (with inverted sign). The graphs of the two functions
may display wide and almost flat regions with values of the order 
of $\epsilon /2$. This may penalize the efficiency in evaluating the integrals, since the stochastic 
sampling of the integration domain would be very weakly driven out of these regions of negligible contribution 
(specially when adopting the small-steps Langevin propagator sensitive to local gradients). For this reason, the choice 
of some value $K > 1$ is preferable since flat regions are avoided and both the details of $f(\bm{x})$ 
where the function is positive-valued, and of $|f(\bm{x})|$  where it is negative-valued, are distributed 
between $f_{+}(\bm{x})$ and $f_{-}(\bm{x})$. Concerning the actual values of $K$ and $\epsilon$, they could be tuned 
by considering their influence on the excursions of $f_{+}(\bm{x})$ and $f_{-}(\bm{x})$, that is, ultimately, their
control on the magnitude of the dimensionless ''potential barriers'' (see Eq. \eqref{eq_pseudo-potential}) to be overtaken 
for an exhaustive exploration of the integration domain. Regardless of such a detailed tuning, it suffices to take $K$ of the order 
of few units, and assign to $\epsilon$ a very small value by checking \emph{a posteriori} that it was actually much smaller than 
the maximum value of $|f(\bm{x})|$ encountered in the explorations of the integration domain.

Notice that the splitting in Eq. \eqref{eq_function_splitting} can be applied \emph{a priori} when 
the behavior of the integrand function within the integration domain is unknown. In fact, even for positive-valued functions one 
would generate $f_{+}(\bm{x}) > f_{-}(\bm{x})$, where $f_{+}(\bm{x})$ is dominant for $\epsilon$ sufficiently small.

As a final remark we shall underline that cases may happen in which $f_{+}(\bm{x})$ and $f_{-}(\bm{x})$
are too close one to the other, so that the global outcome is much smaller (in absolute 
value) compared to them when taking the difference; in these cases it is 
required to reach a very high accuracy on both  $f_{+}(\bm{x})$ and $f_{-}(\bm{x})$, 
with consequent lengthening of the calculation.

\section{Computational issues}\label{sec3}
The methodology outlined in the previous section 
has been implemented in the new JEMDI library, which is distributed under the 
GPL v2.0 license \cite{JEMDI}. The core library files are written in the C++ 
language. The choice followed by the necessity of writing a code that can be easily modified, adapted, and included in 
other algorithms, all issues that can be straightforwardly handled using an object-oriented programming paradigm.
Proper wrappers have been implemented to link the library with C and FORTRAN codes also.
Moreover, the algorithm is parallelized under the message passing interface (MPI) paradigm \cite{MPIa,MPIb}, giving the possibility to run
on computer clusters. The actual parallelization scheme is quite simple: the total number of trajectories is divided 
among the processes allocated for the calculation. This scheme is nearly embarrassingly parallel, thus the computation 
time scales down linearly with the number of processes.
The setup of a calculation may require the intervention of the user really on many parameters. To simplify the 
usage we have set a default behavior of our library, leaving to the expert user the possibility to intervene directly on parameters 
optimization and make personal developments of the code. For technical details and usage examples we address the reader to the 
documentary file \cite{JEMDI}.
To value the tests presented in the next section,
in the Supplementary Material we describe how the method is presently implemented in JEMDI (choice of the 
random numbers generator, criteria for the automatic selection of parameters, internal checks of consistency).
The following algorithm box summarizes the main steps of the procedure.
\begin{algorithm}
\caption{Jarzinsky's Equality Multi-Dimensional Integration}
\begin{algorithmic}
\REQUIRE $f(\bm{x})$, $I$, $N_{tr,b}$, $M$, $N_{\rm steps}$
\ENSURE $E = \int_I d\bm{x} f(\bm{x})$, $\delta E_{\rm stat}$
\STATE{$V_I\leftarrow$ volume of the hyper-rectangle $I$}
\STATE{$N_{tr}\leftarrow N_{tr,b} * M$}
\STATE{$\delta\hat{t}=1/N_{\rm steps}$}
\STATE{$k\leftarrow 1$}
\FOR {$i=1$ to $N_{tr}$}
\STATE {Unbiased random drawing of a starting point $\bm{x}_0 \in I$}
\STATE {Init trajectory work $w_i\leftarrow 0$}
\FOR {$s=1$ to $N_{\rm steps}$}
\STATE{$\hat{t}_s\leftarrow s*\delta\hat{t}$}
\STATE{$\dot{\bm{\Lambda}}(\hat{t}_{s-1})\leftarrow
\displaystyle{
  \frac{d\bm{\Lambda}(\hat{t})}{d\hat{t}}
  {}_{\left\lvert{}_{{\scriptsize \hat{t}=\hat{t}_{s-1}}}\right.}}
$}
\STATE{Function morphing $\bm{\Lambda}(\hat{t}_s) \leftarrow \bm{\Lambda}(\hat{t}_{s-1})+\delta\hat{t}*\dot{\bm{\Lambda}}(\hat{t}_{s-1})$}
\STATE{Propagation (at fixed $\bm{\Lambda}(\hat{t}_s)$)  $\bm{x}(\hat{t}_{s-1})\rightarrow \bm{x}(\hat{t}_s)$, \emph{via} IS-MC or Langevin (eq. 6) propagator}
\STATE{Accumulate work $w_i\leftarrow w_i + \delta\hat{t} * \dot{\bm{\Lambda}}(\hat{t}_{s-1}) \cdot 
\displaystyle{\frac{\partial u(\bm{x},{\bm{\Lambda}})}{\partial \bm{\Lambda}}}_
{\left\lvert{
{\scriptsize \begin{array}{c}\bm{x}=\bm{x}(\hat{t}_{s-1})\\\bm{\Lambda}=\bm{\Lambda}(\hat{t}_{s-1})\end{array}}
}\right.}$}
\ENDFOR
\IF {($N_{tr,b}$ trajectories have been carried out)}
\STATE{Evaluate the partial morphing factor $\Phi_k\leftarrow \sum_j\exp{(-w_j)} / N_{tr,b}$, with $j$ running over 
the latter $N_{tr,b}$ trajectories}
\STATE{$k\leftarrow k+1$}
\STATE{Reinitialize the seed of the pseudo-random numbers generator}
\ENDIF
\ENDFOR
\STATE{Estimate the morphing factor $\Phi\leftarrow\sum_{i=1}^{N_{tr}}\exp{(-w_i)}/N_{tr}$ (eq. 7)}
\STATE{$E\leftarrow V_I*\Phi$}
\STATE{Calculate standard deviation $\sigma_E\leftarrow V_I * \displaystyle{\sqrt{\frac{\sum_{k=1}^M(\Phi_k-\Phi)^2}{M(M-1)}}}$}
\IF{{\bf not} $\sigma_E / E << 1$}
\PRINT{Warning: the calculation shold be repeated with larger $N_{tr}$ or $N_{\rm steps}$.}
\ENDIF
\RETURN {$E$, $\delta E_{\rm stat}\leftarrow \sigma_E$}
\end{algorithmic}
\end{algorithm}
\section{Numerical tests and guidelines}\label{sec4}

For sake of notation, from now on $E$ will replace $E(\bL)$.
To have a reference value of the integral to be taken as exact, we focus on functions
built as products $f(\bm{x})=\Pi_{i=1}^{N/3} \phi(x_{1,i}, x_{2,i}, x_{3,i})$ where
the index $i$ runs from 1 to $N/3$ terns of variables; all factors are taken of equal type for simplicity.
Then we choose the same integration extrema for all corresponding variables in the terns. In this case
$E = E_3^{N/3}$ where $E_3=\int_{a_1}^{b_1} dx_1 \int_{a_2}^{b_2} dx_2 \int_{a_3}^{b_3} dx_3 \, \phi(x_1,x_2,x_3)$
could be evaluated with the highest accuracy by using the DQAND routine from IMSL$^{\textregistered}$ library \cite{IMSL} and/or
with the tools of Mathematica$^{\textregistered}$ package \cite{Mathematica}; in the latter case also an estimate of the error is available.
Tests have been made with functions of $N = 15, 30, 60, 90$ variables \cite{processors}. 
Preliminary explorations on several smooth functions and differently extended integration intervals for the variables 
gave excellent outcomes, hence we shall present here directly the case of functions 
reputed to be hardly integrable by standard means (see below).

We start from a positive-valued function generated as products of
\begin{equation}\label{eq_positive_function}
\phi_A(x_{1,i}, x_{2,i}, x_{3,i}) = e^{-10\cos{\left(2x_{1,i} - 0.5x_{2,i}^3 + 3x_{3,i}\right)}
-5.0 \cos^2{\left(4x_{1,i}^2+8x_{2,i}+2x_{3,i}\right)}}
\end{equation}
The specific form of the exponent has been sorted out just adding trigonometric functions, powers and numerical coefficients
''without thinking'' in doing it. Sectioned contour profiles of such a building block are presented in the
Supplementary Material, revealing a rich morphology made of smooth regions and peaked parts. 
A first integration is done on the hyper-cube with integration extrema -3 and 3 for all variables.

In Figure 1 we show the results by running JEMDI with the IS-MC propagator. For simplicity we fix here
the same maximum length for moves in all dimensions, that is 
$\delta_{{\rm max},i} = \delta_{\rm max}$ for all $i$. 
Calculations refer to equal total number of trajectories
($N_{tr} = 5000$), of equal length ($N_{\rm steps} = 10^5$), partitioned into the same number of blocks to evaluate
the uncertainty $\delta E_{\rm stat} = \sigma_E$ from Eq. \eqref{eq_sigmaE}
($M = 50$, so that $N_{tr,b}= 100$ trajectories per block). 
Horizontal lines give the exact values of the 
integrals, $E^\ast$, while bullets indicate the outcomes $E$ for different values of $\delta_{\rm max}$ with error bars taken as
$\pm \delta E_{\rm stat}$. The numbers associated to each estimate give the average percentage of accepted moves,
$\overline{\%_{acc}}$, here intended over the whole ensemble of paths and over the whole morphing schedule. 
A quick survey indicates that the outcome is satisfactory. Notice that the best results (accurate estimate and small error
bar) are obtained for $\delta_{\rm max}$ values such that the {\em average} acceptance of moves is around 50\% (at least, not 
below 30\%) in accord with the criterion proposed in section \ref{sec2sub4} to guess if the integration domain is
efficiently explored along the morphing. This poses a somehow empirical rule for an automatic selection of the optimal 
values $\delta_{{\rm max},i}$; such a criterion is implemented in JEMDI (see the Supplementary Material for details).

For comparison we have attempted integrations with the standard IS-MC method applied directly to sampling from the
distribution proprortional to
$f(\bm{x})$. For each of 50 runs we have either generated, a) a single Markov chain with a number of steps equal to
$N_{tr,b} \times N_{\rm steps}$, or b) $N_{tr,b}$ chains with $N_{\rm steps}$ each (in both cases the total number of 
steps is equal to that made in a non-equilibrium calculation employing the JE). 
Also, $\delta_{\rm max}$ was optimized to get average acceptance 
percentage of moves close to 50\%. A test was made with a total number of moves equal to $10^7$, to be compared with 
$N_{tr,b} = 100$ and $N_{\rm steps} = 10^5$
in Figure 1. In both cases a) and b) the outcomes were below the exact values by orders of magnitude, which
means a failure of the standard IS-MC integration. Just to mention, for modality a) the average outcomes with $M=50$ 
repetitions for $N = 15, 30, 60, 90$ were respectively $1.3 \times 10^{26}$, 
$5.1 \times 10^{45}$, $5.0 \times 10^{67}$ and $4.2 \times 10^{76}$, incomparable
with the exact values; even worse the outcomes for modality b). For completeness we also tried integration with the basic 
sample-mean MC, that is with unbiased drawing of $10^7$ points in the integration domain: outcomes were even worse. 
Overall, such analysis indicates that our test-function can be classified as hardly integrable, 
and reveals a total failure of the standard MC route while the morphing schedule gives acceptable results
under similar computational cost. The qualitative statement "hardly integrable" stems on the fact that this benchmark
is much more stringent than frequently chosen testing-functions. For example, we have checked that standard 
IS-MC (without morphing) could integrate with excellent accuracy all the testing-functions of ref. \cite{cools07} 
(that is the ${\cal{C}}^0$ and Gaussian Genz's functions, and the Sobol'-Asotsky function, all built as product of 
100 monodimensional factors), but we have seen that it fails here, while the morphing route gives acceptable results. 

Let us turn back to the inaccuracy of the integration when $\delta_{\rm max}$ exceeds a critical value. This makes $E$ dropping below 
$E^\ast$ and stabilizing on a plateau; correspondingly, the average percentage of accepted moves drops to a constant value 
of the order of few units. The rationale of this behaviour lays in the fact that during the morphing a trajectory much probably falls into 
some ''deep potential well'' corresponding to a high function peak, and stays entrapped in it since subsequent IS-MC
moves are likely almost totally rejected if $\delta_{\rm max}$ is so large that a jump would try to take the trajectory 
out of the well (even accounting for possible reflections at boundaries) towards locations of much ''higher potential''.
Some insights on these features are given in the Supplementary Material.
Qualitatively this explains both the low global acceptance of moves (mainly accumulated at the initial stage 
of the morphing when the potential wells are not yet pronounced) and the underestimation of the
integral since some of the peaks may be not visited during the morphing. The small error 
bars are clearly misleading in this case, but consistent with the fact that the integration appears to be  
precise although limited to the visited peaks. Anyway, the low average percentage of accepted moves should warn the user about 
such a pathological situation. 
For $\delta_{\rm max}$ optimized to yield the acceptance of moves
closest to 50\%, in Table 1 we present the outcomes by varying independently the number of trajectories and their length. 
The percentage errors refer to the actual deviations from the exact value. Indication of the execution time is also given. 
Notice how, even in the case of 90 variables, such a function could be integrated with an error less than 
5\% in a reasonable time, while the standard IS-MC counterpart (without morphing) completely fails.

Summarizing, the quality of a calculation with the IS-MC propagator must be valued looking at {\em both} indicators
$\overline{\%_{acc}}$ and $\delta E_{\rm stat}/E$: the former should be above, say, 30\% (we recall that the 
$\delta_{{\rm max},i}$ are automatically optimized in a JEMDI run to bring $\overline{\%_{acc}}$ as close as possible
to 50\%), and the latter must result $\ll 1$ for the systematic error being likely negligible and the 
outcome meaningful (see discussion in section \ref{sec2sub4}). 
On the other hand, there may be situations where both indicators pass the checks but the result is inaccurate:
in all cases it is good practice to assess convergence versus the increasing of $N_{tr}$ and $N_{\rm steps}$.

The same typology of calculations on the same function has been performed by applying Langevin exploration of the integration 
domain, with a scalar (isotropic) "diffusion coefficient" $D$ taken point-independent and constant during morphing. 
The results are presented in Figure 2, where $D$ is varied. 
Notice how the estimated integral reaches a maximum versus increasing $D$, and that 
the same trend (increasing-decreasing) is found for the error bars. 
These features are evident for the cases $N = 60, 90$. Plateaus far from the optimal $D$ value can be 
qualitatively explained as follows.
At low $D$, in the ''timescale'' of the morphing the exploration of the integration domain is limited in the neighborhood 
of the starting point; this situation corresponds, by analogy, to the case of fast-switching 
protocols in the physical context of steered transformations where the system's lag requires a very extended set of runs to reach
accurate outcomes \cite{POH10235}. In the opposite limit of large $D$, the strong damping (note that $D$ tunes the 
magnitude of the ''deterministic force'' in Eq. \eqref{eq_langevin}) drags a trajectory to a function 
maximum, which is very localized for the tightly peaked function here considered (see contour sections in the Supplementary Material). 
In this case, for the 5000 trajectories there might be a relevant 
fraction of peaked regions not visited and hence missing as a matter of fact.
This expectation is confirmed by choosing a random initial point $\bm{x}_0$ and monitoring the Euclidean displacement
$\Delta L = || \bm{x} - \bm{x}_0 ||$ during the morphing. The profiles are presented in
Figure 3 for three values of $D$. The optimal value (the one yielding the best estimate of the integral) is 
$D = 0.30$, which gives a smooth increasing of $\Delta L$ corresponding to departure from the initial point 
with a settling only in the final stage of the morphing. For smaller $D$ there is littler exploration of the domain, while for much 
larger $D$ the trajectory rapidly reaches a plateau which must 
correspond to a fall into a potential well (a peak of the function). In this latter case, a large $D$ value in 
the Langevin equation enhances the magnitude of both the stochastic contribution to the
pseudo-force (notice the magnification of fluctuations amplitude) and the deterministic component which drags the trajectory into
a potential well. In this case, for different values of $D$, different wells seem to be reached even starting 
from the same $\bm{x}_0$ 
  
The global message is that the IS-MC route is {\em a priori }preferable for several reasons: 
{\em i)} it does not require the demanding computation 
of ''forces'', {\em ii)} the empirical criterion based on the average acceptance of moves close to 50\% allows one to parametrize
the IS-MC propagator, {\em iii)} the IS-MC gave better results in all our tests. In particular, the use of our JEMDI 
routine with IS-MC propagation is recommended unless a well educated guess 
from an {\em a priori} knowledge of the function details might address to a proper choice of $D$ (or even to exploit the subtle 
point-dependent tuning, anisotropy, and modulation along morphing in the full diffusion matrix entering Eq. \eqref{eq_langevin}).   

For the same function we have also performed successful tests of integration on hyper-rectangles of very different 
extension on the various dimensions. For example, for the variables $x_{1,i}$ ranging between $-5$ and $+5$, 
$x_{2,i}$ between $-0.002$ and $-0.001$, and $x_{3,i}$ from $1$ and $100$, the outcome with $N = 90$ variables
was $(1.66 \pm 0.36) \times 10^{86}$ which well compares 
with $(1.79 \pm 0.79) \times 10^{86}$ obtained from the integration of the 3-variables building block with Mathematica
($750 \pm 11$); this calculation was done with $N_{tr} = 5000$ trajectories partitioned 
into $M = 50$ blocks to evaluate the uncertainty, $N_{\rm steps} = 10^5$ using IS-MC propagator
with automatic optimization of $\delta_{{\rm max},i}$. Computational time was 6 minutes for a parallel run with the 12-processors 
machine \cite{processors}.

Then we tried the integration of a function with many sign changes according to the splitting in equations 
\eqref{eq_function_splitting}-\eqref{eq_function_Phi}.
Like in the previous case study, the function was built as product of 3-variables building blocks defined as
\begin{eqnarray}\label{eq_changesign_function}
\nonumber
\phi_B(x_{1,i}, x_{2,i}, x_{3,i}) &=& 
-e^{-10\sin{\left( -0.3 x_{1,i}^2 + 4 x_{2,i} + 0.5 x_{3,i}^3 \right)}}+ \\
&&+\phi_A(x_{1,i}, x_{2,i}, x_{3,i})
\end{eqnarray} 
where $\phi_A(x_{1,i},x_{2,i},x_{3,i})$ is still that given in Eq. \eqref{eq_positive_function}.
A single calculation on the hyper-cube of $N = 30$ variables ranging from $-3$ to $+3$ has been done by simulating a 
typical user's situation: automatic selection of $\delta_{{\rm max},i}$, use of default value for the parameter $\epsilon$ (see 
Supplementary Material), and choice of reasonable trial values 
$N_{tr} = 5000$, $M =50$ blocks to evaluate the uncertainty, $N_{\rm steps} = 10^5$, 
for a first run. The outcome was $(-2.8 \pm 10.8) \times 10^{56}$ versus the exact value $2.53 \times 10^{56}$ obtained by 
integration of the 3-variables block with DQAND. The large uncertainty suggests to increase the number of trajectories
and their length; repetition with $N_{tr} = 50000$ (still with $M = 50$) 
and $N_{\rm steps} = 10^6$ gave the satisfactory result $(2.2 \pm 1.8) \times 10^{56}$.
Computational time was 4.6 hours for a parallel run with the 12-processors machine \cite{processors}.   

As final test in our hierarchy of case-studies we considered a function that presents sign changes inside the integration domain 
and singularities at the boundaries, but whose primitive has finite limit on those locations. Preliminary checks revealed that
the method is unable to integrate functions with singular points of divergence {\em inside} the integration domain; on the contrary, 
divergences at boundaries are not problematic if the primitive function is bounded. Again, we selected a function of three 
variables to produce the $N$-dimensional function by means of multiplication of the building block, which is
\begin{equation}\label{eq_logarithmic}
\phi_C(x_{1,i},x_{2,i},x_{3,i})=-\phi_A(x_{1,i},x_{2,i},x_{3,i}) \times \ln(x_{1,i},x_{2,i},x_{3,i})
\end{equation}
Calculations were done for $N$ = 15 and 30 variables with integration ranges 
$x_{1,i} \in [0,1]$, $x_{2,i} \in [0,2]$, and  $x_{3,i} \in [0,3]$.
The exact result in 3 variables obtained with DQAND is 4595.90. Satisfactory convergence of the integration 
has been obtained with $N_{tr} = 50000$, partitioned into $M = 50$ blocks to evaluate the uncertainty, 
$N_{\rm steps} = 10^6$, and by applying
$K = 2$ and $\epsilon = 10^{-5}$ for the parameters discussed in section \ref{sec2sub6}. 
The outcome in 15 variables was $(2.0 \pm 0.1) \times 10^{18}$ versus the exact value of $2.05 \times 10^{18}$, while
in 30 variables we obtained $(4.1 \pm 1.5)\times 10^{36}$ versus the exact value of $4.20 \times 10^{36}$. 
Computational times were respectively 2.6 and 4.7 hours for parallel runs with the 12-processors machine \cite{processors}.
These results still confirm that the method is able to integrate problematic and complex functions without particular problems.
\section{Outlines and perspectives}\label{sec5}
The main purpose of this work was to bring the computational essence of Jarzynski's equality to the pure
numerical context of multivariate integration devoid of any physical trait. Before indicating some
lines of investigation, we remark that here we have presented the very basic Jarzynski's strategy, as it was presented in the
early 1997 article \cite{JAR5018}. We have outlined the main
ideas and presented the basic formulation presently implemented in our JEMDI C++ routine, which revealed to be
high-performing in a series of tests. In particular, our new approach where IS-MC moves are combined with underlying function 
morphing, offers a chance to evaluate multidimensional integrals in an acceptable computational time.
There may be critical situations (depending on the number of variables, features of the integrand, extensions of the integration 
domain) where the convergence rate is still too low, but the present strategy is expected to be much better 
performing than the standard counterpart. Not last, the tool provides the statistical uncertainty 
on the outcome and the criterion to guess if the bias error is negligible; this information is essential to
judge (and control) the reliability of the result.

There is still plenty of room for developments, mainly at the following three levels 
not yet explored: a) selection of a good reference state from which the integrand morphing develops, b) optimization of the 
Markovian propagator, c) tuning of the morphing protocol (i.e., use of non-linear and multi-parameter growth 
of the pseudo-potential) and devising methods to minimize "dissipation". 

The first point should be, in our feeling, the most effective item towards large improvements. Notice that, from the AIS perspective,
adopting the flat reference state corresponds nothing but to adopt the simplest starting distribution which allows to
get independent initial draws and which fully satifies (regardless of the integrand function) 
Neal's indications: "easier to sample from, and which is broad enough to encompass all potential modes" 
(quotation taken from sec. 9 of ref. \cite{Neal01}).

Concerning the point b), we recall that the exploitation of Langevin dynamics deserves consideration since it can be 
a mean for efficient exploration of the integration domain if intuition can address to a proper tailoring of the point-dependent 
(and possibly also progression-dependent) diffusion matrix. This opens a wide landscape to explore; an interested investigator can 
stand on the huge literature about single-molecule dynamics in the overdamped (diffusive) regime of motion with a configuration-dependent 
friction matrix able to affect the pattern of stochastic trajectories giving rise, for example, to saddle-point avoidance phenomena in the
multidimensional pseudo-potential landscape (the logarithm of the positive-valued integrand function) \cite{Moro98}. 
Still concerning the setting of
MC moves, a promising route coming from non-equilibrium simulations of molecular systems under steered transformations 
is that proposed by Chelli \cite{Chelli12} who combined the "configurational freezing" scheme \cite{Nicolini11} with the 
"preferential sampling" approach. The idea is that in steered physical systems, energy dissipation is mainly due to fluctuations 
close to the "hot region" where the external intervent takes place. This led to conceive a scheme where particles are moved within 
a mobility area (roughly, a solvation shell) which encloses the hot region, with a rule to select with preference the particles 
which are closer to the hot region; particles moves are made in the way that the resulting chain is Markovian and detailed balance is guaranteed, as 
required for the applicability of work fluctuations theorems like Jarzynski's equality here treated. Model cases treated in ref. \cite{Chelli12}
were alchemical transformations of a water molecule into a methane molecule in the solvation environment, and 
formation of molecular dimers with solvation. However, how to transfer such a concept
to multidimensional integration is a challenging target: by analogy, we guess that the "mobility region", and the "hot region" within it, 
should be subsets of the whole variables $\bm{x}$ to be determined through a sensitivity-like analysis applied to the pseudo-potential
$u(\bm{x},\bL)$ (in fact, $\delta\bm{x} \cdot \partial u(\bm{x},\bL) / \partial \bm{x}$ is the analogous, in stochastic thermodynamics,
of the the infinitesimal amount of energy exchanged as heat).

About the item c), a starting point could be the work of Schmiedl and Seifert \cite{Seifert07} on the construction of optimal 
protocols, based on the criterion to minimize the average work performed along the non-equilibrium trasformations. 
On physical grounds (Second Principle of thermodynamics 
for isothermal systems at the nanoscale), the average work is higher than the free-energy-difference by an amount that corresponds to the
energy which is dissipated, on average, in driving the transformation. It can be demonstrated that such a dissipation is linked to the spread
of work values and hence, ultimately, to the precision and bias of the outcome when Jarzynski's estimator (Eq. \eqref{eq_estJARZ})
is applied on a finite number 
of realizations. All considerations made for the physical problems can be transferred to the abstract context of multidimensional integration.
A further suggestion to improve the basic route has been proposed by Vaikuntanathan and Jarzynski \cite{Vaiku08}. In what they called "escorted"
transformations, an artificial flow-field of the form $u(\bm{x},\bL)$ (our notation) which "suitably" couples stochastic
variables and controlled parameters, is added to bias the trajectories in the way to minimize (even up to let vanish) the dissipation. 
A reformulation of the JE has been derived by the authors to account for the presence of such a flow-field (see eqs 10, 13 and 14 of the 
cited work). As the authors argue, trial and error experience could lead to contruct
optimal schedules for classes of physical systems; we say that the same would hold also for "classes" (to be properly defined) of integrand 
functions. At last we mention the interesting idea to sample the paths (the trajectories) according to their weight in the exponential
average (see section \ref{sec2sub4}); in the method named Single-Ensemble nonequilibrium Path-Sampling (SEPS) \cite{Ytreberg04}, 
a properly biased sequence of paths is generated using the work as the variable in a Metropolis MC scheme. All these ideas have been tested on
low-dimensional cases, mostly uni- or bi-dimensional, but their application to high-dimensional cases would need to face the 
formidable problem of setting some key-ingredients case by case:
optimal protocol, optimal flow-field, optimal bias function of the low-work wing of the work distribution function. 
On the contrary, our basic implementation 
of Jarzynski's equality has the merit to be directly applicable with the only need, as usual in computational practices, 
to check convergence on the outcomes.           

Finally we like to mention our recent extension of the strategy here presented to the evaluation of nested sums 
over a large number of indexes with positive/negative 
addends (a calculation impossible to tackle by exhaustive evaluation of each addend) \cite{ZFsums}. 
This can be seen as the discrete counterpart of the multidimensional integration.
In this case, the efficiency of the addends morphing (still in
combination with the JE) can be quantified, and appreciated, by looking at the incredibly small ratio between the number of
required addends evalutations versus the total number of addends.
\\ \\
\noindent
{\bf Acknowledgments}
Calculations were run on the HPC hardware of the "Centro di Chimica Computazionale di Padova" (C3P) hosted at the
Department of Chemistry of the University of Padova.


\begin{thebibliography}{00}
\bibitem{notex3}
Turning for a while to the AIS conterpart of the JE, we remark that a similar conclusion was already 
formulated by Neal who stated that "[...] nightmare scenarios in which wrong results are obtained 
without there being any indication of a problem are possible [...] this can occur when the distribution
of the importance weights has a heavy upward tail that is not apparent from the data collected"
(quotation taken from sec. 9 of ref. \cite{Neal01}; Neal's weights correspond to
the terms $\exp(- w_i)$, hence he refers to our low-work wing).
\end{thebibliography}

\clearpage

\clearpage
\begin{table}[ht]
\centering
\begin{tabular}{c c c c c c}
N var & $N_{tr}$ & $N_{\rm steps}$ & $\delta_{\rm max}$ & \% error & time / s\cr
\hline
15 & 5000  & $10^5$ & $3.981\cdot10^{-2}$ & 0.66  & 70 \cr
   &      & $10^6$ &                     & -0.17 & 606\cr
   & 50000 & $10^5$ &                     & -0.05 & 597\cr
   &      & $10^6$ &                     & 0.03  & 5376\cr
\hline
30 & 5000  & $10^5$ & $3.981\cdot10^{-2}$ & 0.54 & 117 \cr
   &      & $10^6$ &                     & -0.41 & 1129 \cr
   & 50000 & $10^5$ &                     & -0.48 & 1022 \cr
   &      & $10^6$ &                     & -0.14 & 10025\cr
\hline
60 & 5000  & $10^5$ & $1.995\cdot10^{-2}$ & 3.05 & 221 \cr
   &      & $10^6$ &                     & -1.43 & 2195 \cr
   & 50000 & $10^5$ &                     & 0.92 &  1998\cr
   &      & $10^6$ &                     & -0.28 & 21634\cr
\hline
90 & 5000  & $10^5$ & $1.995\cdot10^{-2}$ &  -14.95& 312 \cr
   &      & $10^6$ &                     & -1.29 & 3076 \cr
   & 50000 & $10^5$ &                     & -4.70 & 3363\cr
   &      & $10^6$ &                     & -0.003 & 30106\cr
\hline
\end{tabular}
\vspace*{1cm}
\caption{Comparison between integral evaluations by varying number and
length of the IS-MC trajectories 
(errors are estimated by partitioning the trajectories into $M =50$ blocks). 
The chosen $\delta_{\rm max}$ is the one that gave the average
acceptance of moves closest to 50\%. Computation times are evaluated
for a parallel run over 16 processors.}
\end{table}

\clearpage

\begin{figure}\centering
\includegraphics[width=\textwidth]{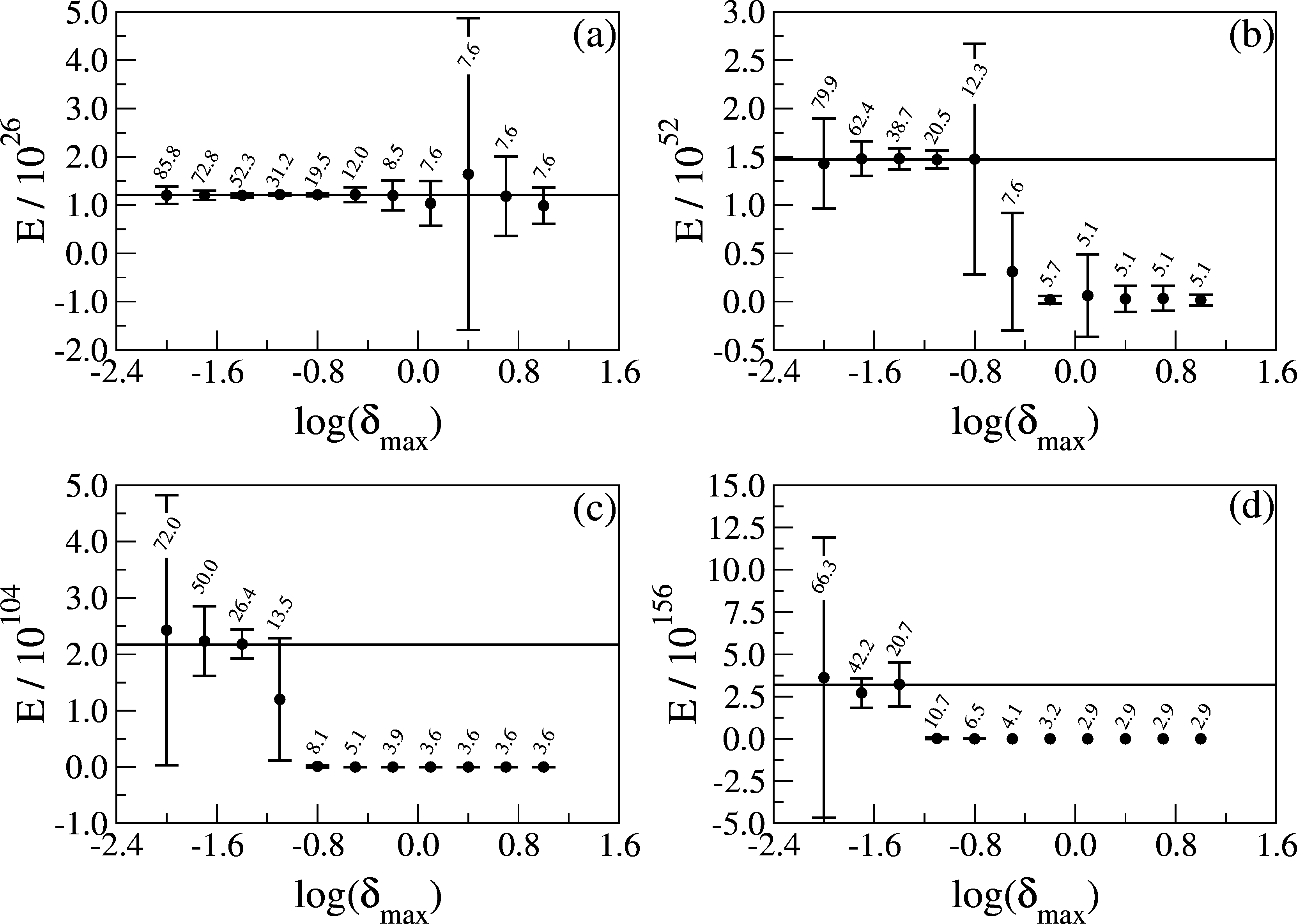}
\caption{
Estimated integral of the test-function generated by building-blocks $\phi_A$ (see text) 
over the domain from -3 to +3 per each variable, with  
(a) 15, (b) 30, (c) 60, and (d) 90 variables; values are given versus the maximum length of IS-MC moves per each dimension, 
$\delta_{\rm max}$. 
Estimates are obtained with 5000 IS-MC trajectories, each of length $10^5$ steps
(partition into $M = 50$ blocks is taken to estimate the errors). 
Horizontal lines show the exact values respectively of  $1.213\cdot10^{26}$, $1.472\cdot10^{52}$,
$2.167\cdot10^{104}$, and $3.189\cdot10^{156}$. Labels on points indicate the average percentage acceptance of moves.
}
\end{figure}

\clearpage

\begin{figure}\centering
\includegraphics[width=\textwidth]{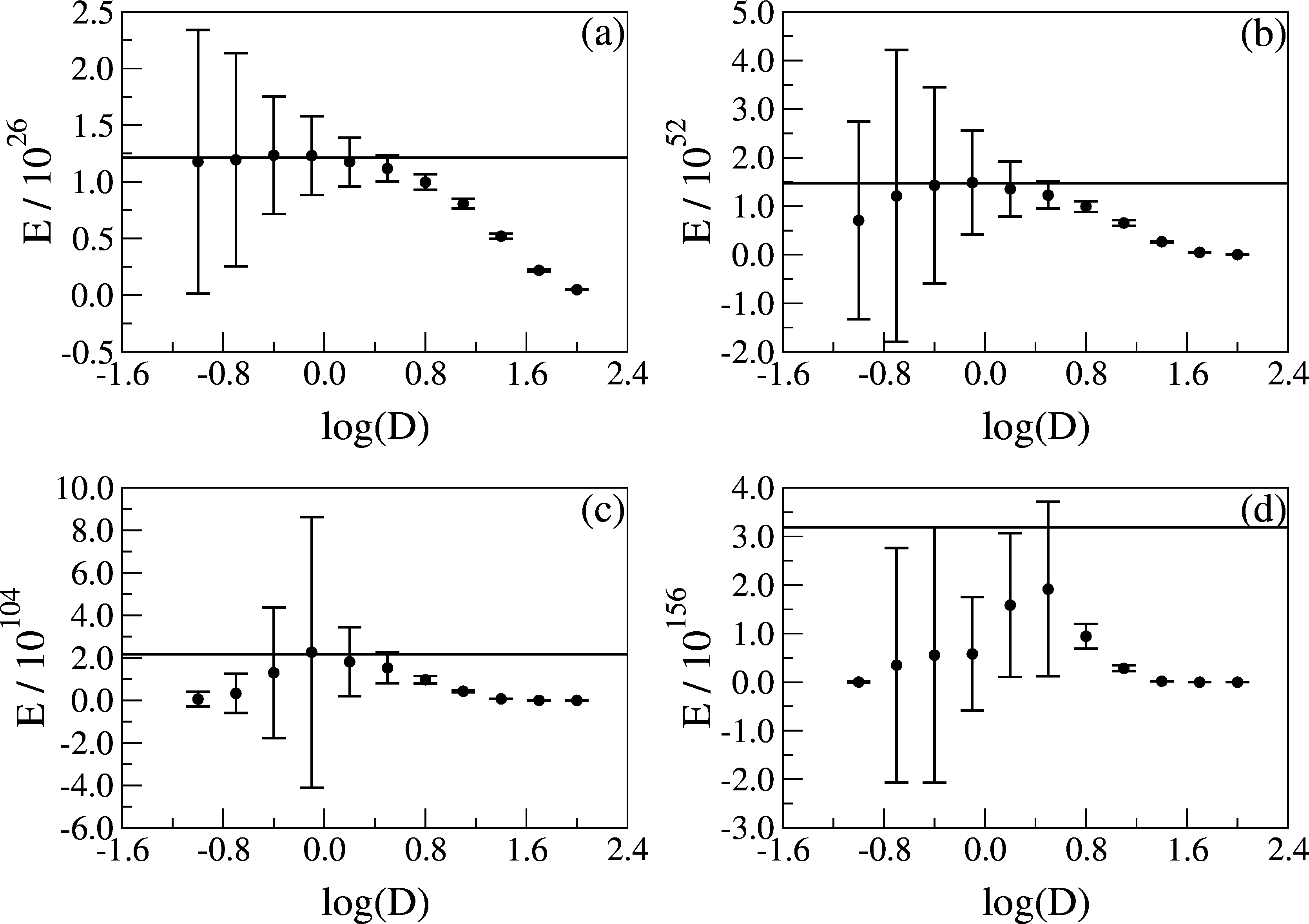}
\caption{
Estimated integral of the test-function generated by building-blocks $\phi_A$ (see text) 
over the domain from -3 to +3 per each variable, with  
(a) 15, (b) 30, (c) 60, and (d) 90 variables; values are displayed versus $D$. 
Estimates are obtained with 5000 Langevin trajectories, each of length $10^5$ steps
(partition into $M = 50$ blocks is taken to estimate the errors). 
Horizontal lines show the exact values respectively of  $1.213\cdot10^{26}$, $1.472\cdot10^{52}$,
$2.167\cdot10^{104}$, and $3.189\cdot10^{156}$.
}
\end{figure}

\clearpage

\begin{figure}\centering
\includegraphics[width=\textwidth]{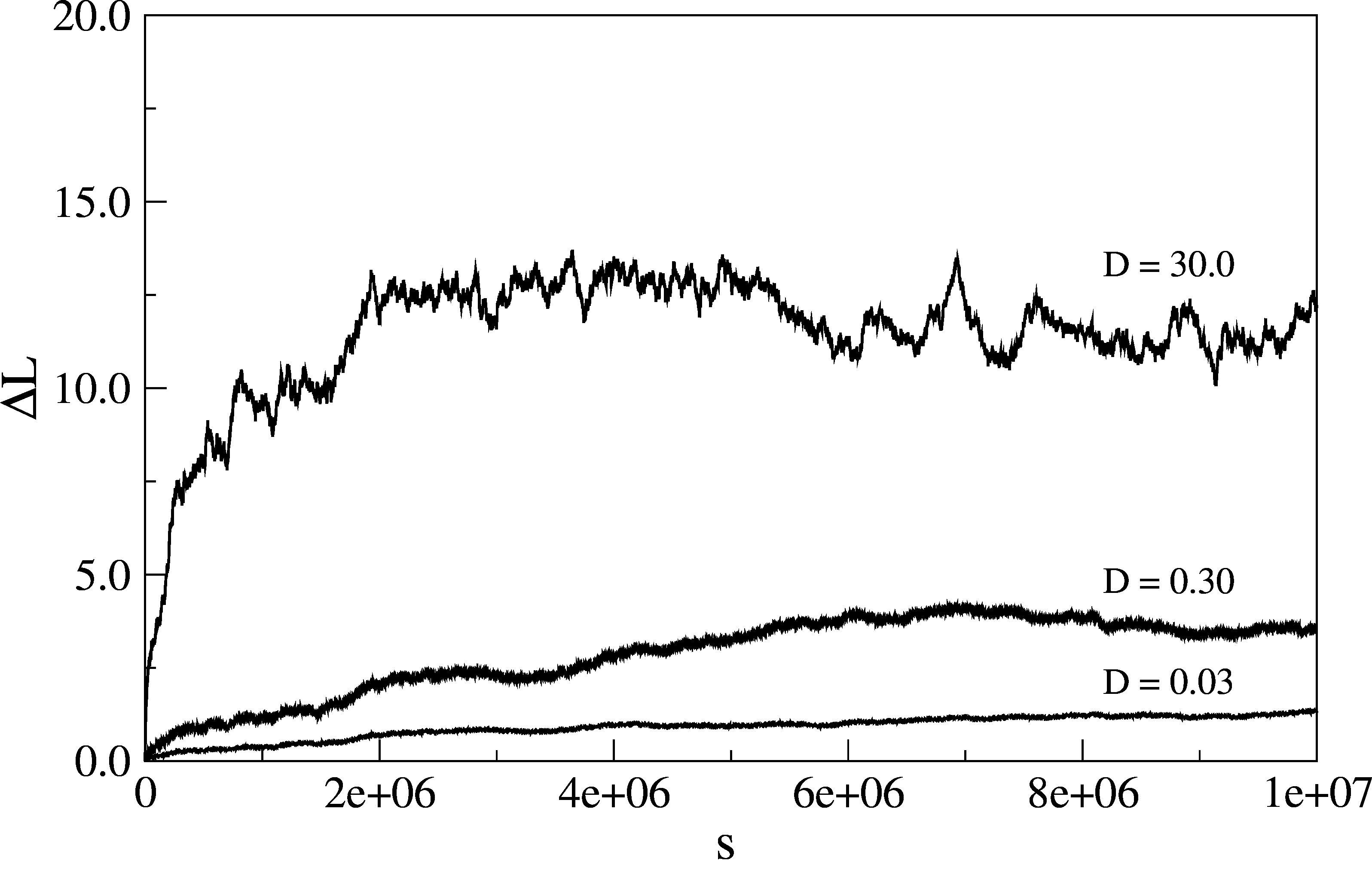}
\caption{
Plot of the Euclidean displacement $\Delta L$ vs. the number of Langevin trajectory steps for the case 
$N = 30$ variables displayed in Figure 2, and for three values of the diffusion
coefficient: $D$ = 0.03, 0.3 , 30.0. $\Delta L$ has been evaluated with respect
to a starting point $\bm{x}_0$ randomly generated only once, and then applied in all the three calculations.
}
\end{figure}

\end{document}